# Designing compensated magnetic states in tetragonal Mn$_3$Ge-based alloys

Yurong You, Guizhou Xu[*], Fang Hu, Yuanyuan Gong, Er Liu, Feng Xu[*]

School of Materials Science and Engineering, Nanjing University of Science and Technology, Nanjing 210094, China

**Abstract**

Magnetic compensated state attracted much interests due to the observed large exchange bias and large coercivity, and its potential applications in the antiferromagnetic spintronics with merit of no stray field. In this work, by *ab initio* calculations with KKR-CPA for the treatment of random substitution, we obtain the complete compensated states in the Ni (Pd, Pt) doped Mn$_3$Ge-based D0$_{22}$-type tetragonal Heusler alloys. We find the total moment change is asymmetric across the compensation point (at ~ x = 0.3) in Mn$_{3-x}$Y$_x$Ge (Y = Ni, Pd, Pt), which is highly conforming to that experimentally observed in Mn$_3$Ga. In addition, an uncommon discontinuous jump is observed across the critical zero-moment point, indicating that some non-trivial properties can emerge at this point. Further electronic analysis for the three compensation compositions reveals large spin polarizations, together with the high Curie temperature of the host Mn$_3$Ge, making them promising candidates for spin transfer torque applications.

***Keywords:*** Tetragonal Heusler alloys, magnetic compensated state, Mn$_3$Ge

[*] *Email:* gzxu@njust.edu.cn (G. Xu), and xufeng@njust.edu.cn (F. Xu).

# 1. Introduction

The tetragonal Heusler compounds, namely a kind of tetragonal distorted variant of the cubic Heusler alloys, have attracted significant research interests these years, due to their novel magnetic structures [1-3], large perpendicular magnetocrystalline anisotropy, and potential applications in spintronics and permanent magnets [4,5]. Tetragonal Heusler alloys are usually ferrimagnetic with at least two magnetic sublattices which are aligned antiparallel or noncollinear to one another [1,2], offering opportunities to continuously tune their magnetic properties for diverse applications. Among them, Mn-based materials draw special attention. For instance, the tetragonal $Mn_3Ga$, on account of its combination of high spin polarization, high Curie temperature, low magnetic moment and relatively low damping, has been widely studied for spin transfer torque (STT) applications [6-8]. $D0_{22}$-type tetragonal $Mn_3Ga$ is a ferrimagnet with a net compensation of 1 $\mu_B$/f.u., and by substitution of the Mn atom, a special compensated magnetic state can be achieved, with the total magnetic moment to be zero while the partial Mn moment nonzero [9]. Magnetic compensated materials can play a vital role in the antiferromagnetic spintronics with the merit of no stray field [10]. Also, they can embrace large exchange bias and large coercivity due to the exchange anisotropy induced by the ferrimagnetic clusters embedded in the compensated host [11]. The exchange bias has reached more than 3 T in the bulk Mn-Pt-Ga compensated states, however, the field to saturate the magnetic moments is still relatively large.

Besides $Mn_3Ga$, $Mn_3Ge$ can also exhibit tetragonal Heusler structure under specific heat treatments [12] and the perpendicular magnetic anisotropy has been demonstrated in its thin film form [13]. If the magnetic states of $D0_{22}$-type $Mn_3Ge$ can be tuned to a compensated state, the large exchange bias effect can be expected like in Mn$_3$Ga, making it be another good candidate for STT and antiferromagnetic spintronics. In this work, we attempt to tune the magnetic properties and find a compensated magnetic state based on Mn$_3$Ge-based tetragonal structure by an *ab initio* study. Considering that Mn$_2$NiGe inclines to form the tetragonal structure instead of the cubic one [14], we choose Ni and isoelectronic Pd, Pt to substitute Mn atoms, respectively. The first-principle calculations show that the electronic structure gradually changes with increasing Ni (Pd, Pt) content, and the compensated states are finally obtained. In addition, we find that the compensated compounds are highly spin-polarized, making them attractive for highly polarized antiferromagnetic applications [9].

**2. Calculation Details**

We perform *ab initio* density functional calculations to search the compensated states in Mn$_{3-x}$Y$_x$Ge (Y = Ni, Pd, Pt) alloys. In order to treat the continuous non-stoichiometric situation, we apply the Korringa-Kohn-Rostoker method combined with the coherent potential approximation (KKR-CPA) [15-17], which is implemented in the AkaiKKR code [18]. KKR is an all-electron method with high speed and high accuracy, and CPA is taken as the most efficient scheme to deal with random systems at present [19]. In the calculation, the exchange correlation effect is treated with generalized gradient approximation (GGA) function [20]. For comparison, we also calculate the stoichiometric Mn$_3$Ge and Mn$_2$YGe with the CASTEP code [21, 22], a pseudopotential method based on plane-wave basis set. It presents similar results with KKR-CPA when the lattice parameters are set to be same.

**3. Results and discussions**

Fig. 1 shows the tetragonal structure of Mn$_3$Ge (a) and Mn$_2$YGe (b). The D0$_{22}$-type tetragonal Mn$_3$Ge, similar to Mn$_3$Ga, can also be seen as a distortion from the cubic Heusler structure, thus the atomic occupation obeys certain rules as in the Heusler alloys[12]. There are two types of Mn locations: one occupies 4d (0, 1/2, 1/4), and is denoted as $Mn^A$ (blue); another occupies 2b (0, 0, 1/2), and is denoted as $Mn^B$ (purple). The main group Ge element occupies 2a (0, 0, 0) site. According to the early neutron scattering experiments [12], the magnetic moments of $Mn^A$ and $Mn^B$ align antiferromagnetically parallel to c-axis, as depicted in Fig. 1a. In the $Mn_2YGe$ alloys, the electron occupation rule [12] determines that Y will prefer to occupy 4d site when it possesses more electrons than Mn, as seen in Fig. 1b. The calculations with CASTEP and KKR-CPA for $Mn_3Ge$ reveal very close results, where the magnetic moments of $Mn^A$ and $Mn^B$ are 1.89 $\mu_B$ and -2.88 $\mu_B$, respectively. The spin density of $Mn_3Ge$ in Fig. 1c reveals that the magnetization mainly concentrates on the Mn atoms, and the non-equivalent Mn atoms bring a ferrimagnetic state (the yellow and blue isosurfaces represent opposite spin directions). The substitution with a third atom not only changes the proportions of $Mn^A$ and $Mn^B$ atoms, but also changes the distribution of their magnetization. The spin density of $Mn_2NiGe$ in Fig. 1d shows that the moment of $Mn^A$ becomes larger indicated by the grown isosurface (yellow one), while that of $Mn^B$ keep nearly unchanged and Ni atom (orange) owns a small moment with the same sign to that of $Mn^B$. The total magnetization is reversal for $Mn_3Ge$ and $Mn_2NiGe$, thus the completely compensated magnetic state is anticipated realizable by gradually replacing the $Mn^A$ atoms with Ni (Pd, Pt) atoms.

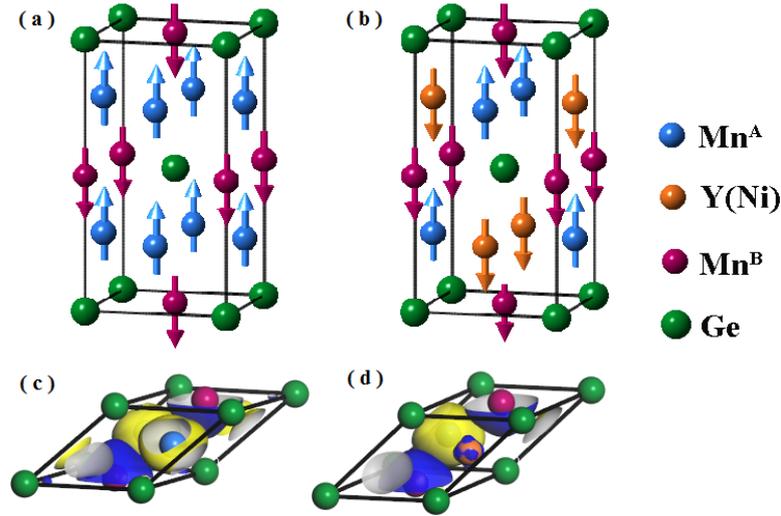

Fig. 1. The $D0_{22}$ tetragonal structure of (a) $Mn_3Ge$ and (b) $Mn_2YGe$, where $Mn^A$ occupies 4d (0, 1/2, 1/4), $Mn^B$ occupies 2b (0, 0, 1/2), Ge occupies 2a (0, 0, 0), Y occupies one of $Mn^A$ site. The electron spin density displayed in the primitive cell of (c) $Mn_3Ge$ and (d) $Mn_2YGe$ (Y=Ni). The yellow isosurface reveals the spin-up magnetization and the blue one reveals the opposite.

Based on the above tetragonal structure, we obtained the optimized lattice parameters of $Mn_3Ge$ with CASTEP calculation, which are a = 3.738Å and c = 6.986Å, close to the reported value [11,23]. With this parameter, we carried out calculations by KKR-CPA for non-stoichiometric $Mn_{3-x}Ni(Pd, Pt)_xGe$ alloys. Fig. 2 summarizes the variation of the magnetization (total and partial) in the whole substitution process. The total magnetic moment is shown in absolute value, while the sign of partial moment represents their relative direction. It is found that all three alloy series present similar variation trend. With increasing Y content, the total magnetic moment first drops quickly in a linear manner, reaching a zero magnetization state at about x = 0.3, then grows in a parabolic way across the zero point. This asymmetric variation of the moment across the zero point highly accords with what has been experimentally observed in $Mn_3Ga$ system [11], proving the validity of our calculation in turn. These results can be a good guidance for future experiments in $Mn_3Ge$ system.

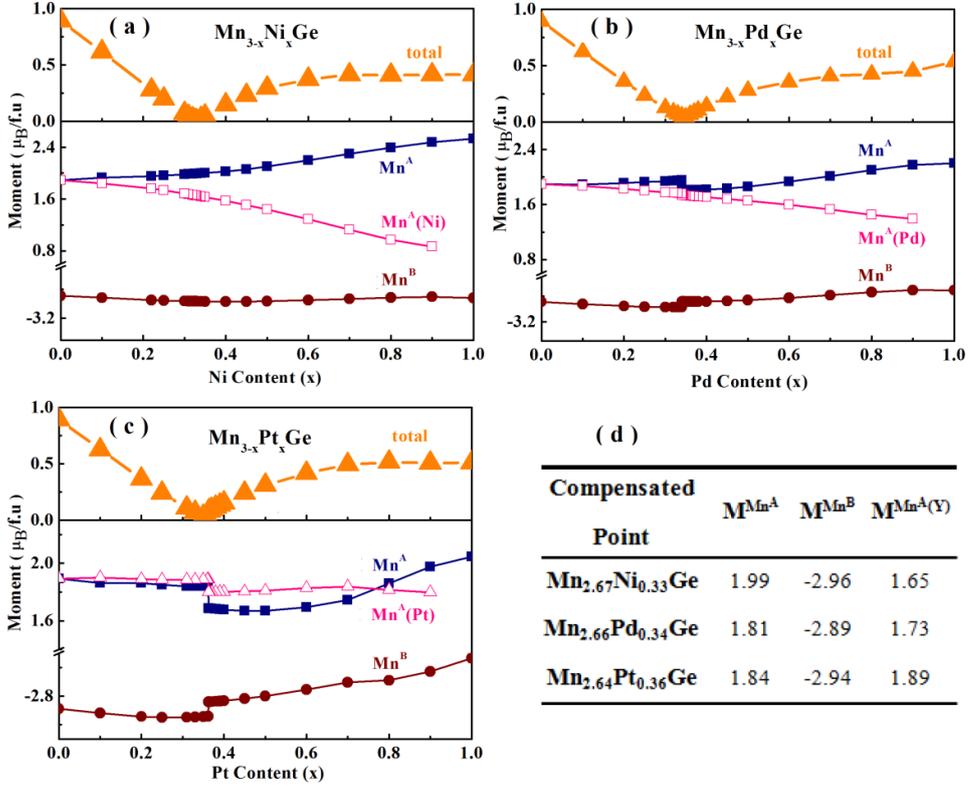

Fig. 2. The calculated total and partial moment of (a) $Mn_{3-x}Ni_xGe$, (b) $Mn_{3-x}Pd_xGe$, (c) $Mn_{3-x}Pt_xGe$. $Mn^A$(Ni/Pd/Pt) represents the Mn atom that is partially substituted with Ni/Pd/Pt atoms. (d) List of exact compositions of the compensation alloys and their Mn moments.

To understand the overall trend, we investigated the atom-resolved moment change. Since Ni, Pd and Pt possess little (less than 0.4 $\mu_B$ and even zero) moments, we mainly inspected the moment variation of Mn atoms, as shown in the lower panel of Fig. 2. Take the case of Mn-Ni-Ge (Fig. 2a) as an example, in the lower-Ni part, the magnetic moments of the two $Mn^A$ atoms and $Mn^B$ atom stay nearly unchanged. Consequently, the decrease of $Mn^A$ population directly causes the linear decrease of the total magnetic moment. The further increase of Ni-content exerts an influence upon the surrounding Mn magnetization: the $Mn^A$ atom that is not substituted shows increase of the moment, while the other $Mn^A$ (partially substituted by Ni) shows opposite trend. Together with the further decrease of $Mn^A$ population, the total moment increases to almost saturation following a parabolic-like climbing. It should be noted that an unusual discontinuous jump was observed across the zero point in the alloy series of Pd and Pt substitution, as shown in Fig. 2b and 2c. This indicates an abnormal transition takes place at this critical zero point, where novel physics or properties can emerge, and they will become our future research topics. In Fig. 2d, we list the exact compositions of the compensated alloys, which are $x_{Ni} = 0.33$, $x_{Pd} = 0.34$ and $x_{Pt} = 0.36$. The average moments of $Mn^A$ (~ 1.8 $\mu_B$) and $Mn^B$ (~ -2.9 $\mu_B$) changes slightly compared to that of $Mn_3Ge$ ($Mn^A$ is 1.89 $\mu_B$, $Mn^B$ is -2.88 $\mu_B$). It again indicates that the compensation state is mainly caused by the atom proportion change.

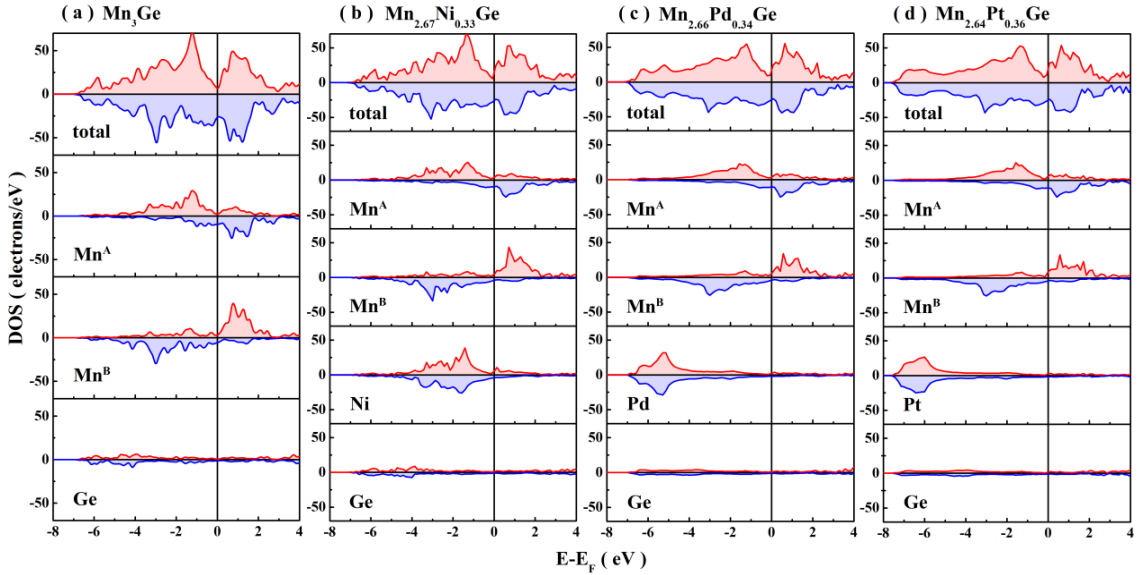

Fig. 3. Total and partial spin polarized density of states (DOS) for $Mn_3Ge$ (a) and the compensated alloys (b, c, d). As $Mn^A$ and $Mn^A(Y)$ possess similar DOS, we only present $Mn^A$ in (b, c, d).

To further inspect the electronic and magnetic properties of above alloys, we have calculated the density of states (DOS) of the compensated compounds, as well as the host alloy $Mn_3Ge$ (Fig. 3). It can be seen that in $Mn_3Ge$, there is a big dip at the Fermi level in the spin up electronic band, bringing spin polarization to be 70.7%, which is comparable to the widely-studied $Mn_3Ga$ alloys [7]. There are two main peaks in the spin-polarized DOS. One is located at about -0.15 eV in the spin up direction, the other is at about -0.25 eV in the spin-down direction, corresponding to the $Mn^A$ and $Mn^B$ bonding states judging from the partial DOS. The overall DOS of the compensated alloys are resembling to that of $Mn_3Ge$, but are modified due to the mixture of Y (Ni/Pd/Pt) with $Mn^A$, leading to broader peaks. The spin polarizations for $Mn_{2.67}Ni_{0.33}Ge$, $Mn_{2.66}Pd_{0.34}Ge$, $Mn_{2.64}PtGe_{0.36}$ are 66.6%, 48.1% and 42.0%, respectively, showing the advantages for the application in the highly-polarized antiferromagnetic spintronics.

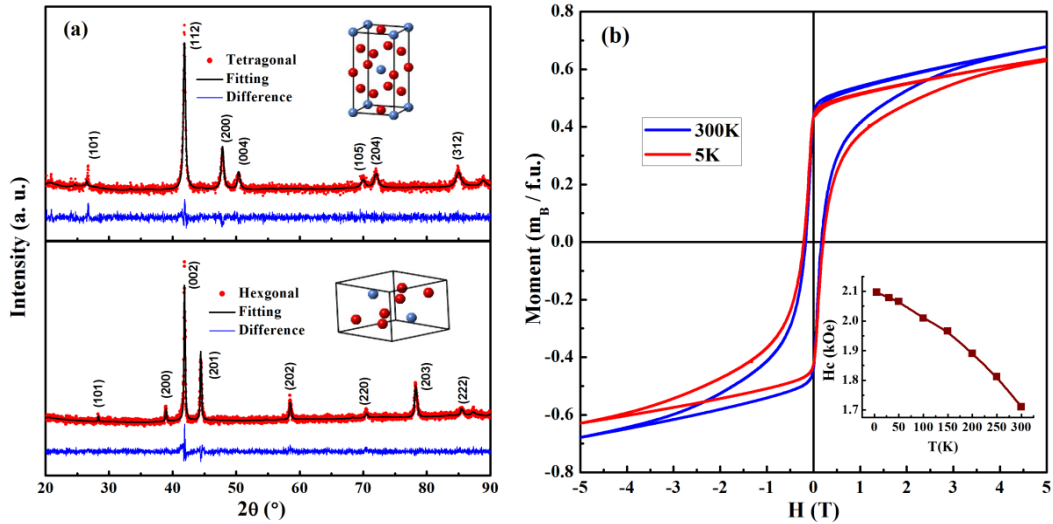

Fig. 4. (a) Powder XRD pattern of the tetragonal $D0_{22}$ structure and hexagonal $D0_{19}$-type $Mn_{3.4}Ge$ alloys. The red line represents the original experimental data, the black

line is the Rietveld fitting line and the blue one is the difference between them. (b) Isothermal magnetization curves (M-H) for tetragonal $Mn_{3.4}Ge$ at 5K and 300K. The inset shows the temperature dependence of the coercivity field $H_c$.

To obtain the compensated alloys, we firstly try to synthesize $Mn_3Ge$ in the tetragonal phase. According to the early report, single phase of $Mn_3Ge$ exists only when the composition is among the range between $Mn_{3.2}Ge$ and $Mn_{3.4}Ge$, since the excess Mn atoms that occupy the site of Ge can help to stabilize a single hexagonal $D0_{19}$ or tetragonal $D0_{22}$ phase [24-26]. We synthesized the high-temperature hexagonal phase by annealing the as-cast $Mn_{3.4}Ge$ samples at 800°C for 5 days before water-quenching, then the tetragonal phase was obtained by annealing the hexagonal samples at 450°C for 2 weeks. As Fig. 4 shows, the refined lattice constants of tetragonal samples are a = 3.803 Å and c = 7.236 Å, and those of hexagonal samples are a = 5.337 Å and c = 4.311 Å. The magnetization measurements of tetragonal samples at 5 K show the magnetic moment of 0.63 $\mu_B$ at 5T, which is a little smaller than the theoretical value of 0.9 $\mu_B$ owing to the non-saturated MH behavior. With temperature increase, the high-field magnetization nearly keeps unchanged, while at 300K, it enhances a bit instead. This indicates the high Curie temperature of $Mn_3Ge$, which is reported as high as 920 K [12]. The M-H curve shows a contraction behavior, resembling to that of the hexagonal $Mn_3Ge$ [27], with a coercivity field $H_c$ in the range of 1.7-2.1 kOe from 300K to 5K. To observe the exchange bias effect in the tetragonal sample, the field cooling from above the Curie temperature is required. The experimental realization of compensated compositions will need further studies in the future.

**4. Conclusions**

In summary, we have achieved to design fully compensated magnetic states in Mn$_3$Ge-based tetragonal structure by *ab initio* calculations. We find the compensation point is at about x = 0.3 in the alloys of Mn$_{3-x}$Y$_x$Ge (Y=Ni, Pd, Pt). The moment change is asymmetric across the compensation point which is highly conforming to the Mn$_3$Ga system, indicating the experimental feasibility based on our calculations. In addition, an uncommon discontinuous jump is observed across the critical point, and it can embrace novel phenomena in experiments. The DOS calculations for the three compensation composition reveal large spin polarization. The preliminary experimental results for Mn$_3$Ge show the realization of single tetragonal phase and its consistency to the calculation results.

**Acknowledgement**

This work is sponsored by National Natural Science Foundation of China (51271093, 51571121, 11604148, 51601092 and 51601093), Fundamental Research Funds for the Central Universities (Grant No: 30920140111010, 30916011345 and 30916011344), Jiangsu Natural Science Foundation for Distinguished Young Scholars (BK20140035), Natural Science Foundation of Jiangsu Province (BK20160833, 20160829 and 20160831). It is also funded by Qing Lan Project, Six Talent Peaks Project in Jiangsu Province, and the Priority Academic Program Development of Jiangsu Higher Education Institutions.

**Reference**

[1] S. Singh, R. Rawat, S. Esakki Muthu, S. W. D. Souza, E. Suard, A. Senyshyn, S. Banik, P. Rajput, S. Bhardwaj, A. M. Awasthi, Rajeev Ranjan, S. Arumugam, D.

L. Schlagel, T. A. Lograsso, A. Chakrabarti, S.R. Barman, Phys. Rev. Lett. 109 (2012) 246601.

[2] O. Meshcheriakova, S. Chadov, A. K. Nayak, U. K. Rößler, J. Kübler, G. André, A. A. Tsirlin, J. Kiss, S. Hausdorf, A. Kalache, W. Schnelle, M. Nicklas, C. Felser, Phys. Rev. Lett. 113 (2014) 087203.

[3] S. Singh, J. Nayak, E. Suard, L. Chapon, A. Senyshyn, V. Petricek, Y. Skourski, M. Nicklas, C. Felser, S. Chadov, Nat. Commun. 7 (2016) 12671.

[4] J. Winterlik, S. Chadov, A. Gupta, V. Alijani, T. Gasi, K. Filsinger, B. Balke, G. H. Fecher, C. A. Jenkins, F. Casper, J. Kübler, G. D. Liu, L. Gao, S. S. P. Parkin, C. Felser, Adv. Mater. 24 (2012) 6283.

[5] T. J. Nummy, S. P. Bennett, T. Cardinal, D. Heiman, Appl. Phys. Lett. 99 (2011) 252506.

[6] B. Balke, G. H. Fecher, J. Winterlik, C. Felser, Appl. Phys. Lett. 90 (2007) 152504.

[7] H. Kurt, K. Rode, M. Venkatesan, P. Stamenov, J. M. D. Coey, Phys. Rev. B 83 (2011) 020405.

[8] J. Winterlik, B. Balke, G. H. Fecher, C. Felser, M. C. M. Alves, F. Bernardi, J. Morais, Phys. Rev. B 77 (2008) 054406.

[9] X. Hu, Adv. Mater. 24 (2012) 294.

[10] T. Jungwirth, X. Marti, P. Wadley, J. Wunderlich, Nat. Nanotech. 11 (2016) 231.

[11] A. K. Nayak, M. Nicklas, S. Chadov, P. Khuntia, C. Shekhar, A. Kalache, M. Baenitz, Y. Skourski, V. K. Guduru, A. Puri, U. Zeitler, J. M. D. Coey, C. Felser, Nat. Mat. 14 (2015) 679.

[12] G. Kadar, E. Kren, Int. J. Magn. 1 (1971) 143.

[13] H. Kurt, N. Baadji, K. Rode, M. Venkatesan, P. Stamenov, S. Sanvito, J. M. D. Coey, Appl. Phys. Lett. 101 (2012) 132410.

[14] H. Z. Luo, F. B. Meng, G. D. Liu, H. Y. Liu, P. Z. Jia, E. K. Liu, W. H. Wang, G. H. Wu, Intermetallics 38 (2013) 139.

[15] H. Katayama, K. Terakura, J. Kanamori, Solid State Commun. 29 (1979) 431.

[16] S. Blugel, H. Akai, R. Zeller, P. H. Dederichs, Phys. Rev. B 35 (1987) 3271.

[17] S. Kaprzyk, A. Bansil, Phys. Rev. B 42 (1990) 7358.

[18] H. Akai, Hyperfine Interact. 68 (1992) 3.

[19] W. H. Butler, Phys. Rev. B 31 (1985) 3260.

[20] J. P. Perdew, J. A. Chevary, S. H. Vosko, K. A. Jackson, M. R. Pederson, D. J. Singh, C. Fiolhais, Phys. Rev. B 46 (1992) 6671.

[21] M. C. Payne, M. P. Teter, D. C. Allan, T. A. Arias, J. D. Joannopoolous, Rev. Mod. Phys. 64 (1992) 1065.

[22] M. D. Segall, P. L. D. Lindan, M. J. Probert, C. J. Pickard, P. J. Hasnip, S. J. Clark, M. C. Payne, J. Phys.: Condens. Matter 14 (2002) 2717.

[23] D. L. Zhang, B. H. Yan, S. C. Wu, J. Kübler, G. Kreiner, S. S. P. Parkin, C. Felser. J. Phys.: Condens. Matter 25 (2013) 206006.

[24] N. Yamada, J. Phys. Soc. Jpn. 59 (1990) 273.

[25] N. Yamada, H. Sakai, H. Mori, T. Ohoyama, Physica B 149 (1988) 311.

[26] A. Kalache, G. Kreiner, S. Ouardi, S. Selle, C. Patzig, T. Höche, C. Felser, APL Mater. 4 (2016) 086113.

[27] J. F. Qian, A. K. Nayak, G. Kreiner, W. Schnelle, C. Felser, J. Phys. D: Appl. Phys. 47 (2014) 305001.